\documentclass[11pt]{article}   	
\usepackage{geometry}                		
\usepackage{graphicx}				
\usepackage{amssymb}

\usepackage{graphicx}

\RequirePackage{amssymb}

\begin{document}

\title{Torsion Effects and LLG Equation.}

\date{}

\maketitle

\begin{center}
\author{Cristine N. Ferreira$^{\dagger a} $, Cresus F. L. Godinho $^{\ddagger b}$ ,  J. A. Helay\"{e}l Neto} $^{* c}$ \\
\end{center}

\begin{center}

$^\dagger$ N\'ucleo de Estudos em F\'{\i}sica,  Instituto Federal de Educa\c{c}\~{a}o,
Ci\^encia e Tecnologia Fluminense,
Rua Dr. Siqueira 273, Campos dos
Goytacazes, 28030-130
RJ, Brazil \\

$^\ddagger$ Grupo de F\' isica Te\'orica, 
Departamento de F\'{\i}sica, Universidade Federal Rural do Rio de Janeiro,
BR 465-07, 23890-971, Serop\'edica, RJ, Brazil \\

$^*$ Centro Brasileiro de Pesquisas F\' isicas (CBPF),
Rua Dr. Xavier Sigaud 150, Urca,
22290-180, Rio de Janeiro, Brazil \\

\end{center}

\begin{abstract}
Based on the non-relativistic regime of the Dirac equation coupled to a torsion pseudo-vector, we study the dynamics of magnetization and how it is affected by the presence of torsion.  We consider that torsion interacting terms in Dirac equation appear in  two ways  one of these is thhrough the covariant derivative considering the spin connection and gauge magnetic field and the other is through a non-minimal spin torsion coupling.
We show within this framework, that it is possible to obtain the most general Landau, Lifshitz and Gilbert (LLG) equation including the torsion effects, where we refer to torsion as a geometric field playing an important role in the spin coupling process.  We show that the torsion terms can give us two important landscapes in  the magnetization dynamics: one of them related with damping and the other related with the screw dislocation that give us a global effect like a helix damping sharped. These terms are responsible for changes in the magnetization precession dynamics. 						
\end{abstract}

\vspace{1 true cm}

 \texttt{$^{a}$crisnfer@iff.edu.br},
 \texttt{$^{b}$crgodinho@ufrrj.br},  \texttt{$^{c}$helayel@cbpf.br}

\newpage

\section{Introduction}

The discovery of the graphene-like systems and topological insulators systems introduced a new dynamic in the applications of the framework of the high energy physics in low energy systems in special condensed matter systems. The fact that these systems can be described by Dirac equations give us new possibilities for theoretical and experimental applications.  In this direction there are some  effects in condensed matter systems still without a full description as the magnetizable systems that we described in this work. In this sense the construction of the theoretical frameworks that can, in some limit, be obtained in low energy systems is the crucial importance to understand  the invariances and interactions in certain limits. One of the important effects that we can study is related with the spin systems. So, in this work we deal with the new framework to study the spin systems considering the Dirac equation in non-relativistic limit with torsion interaction\cite{Dyrdal}. 
Spin systems are generally connected with magnetic systems.  It is well known that spin angular momentum is an intrinsic property of quantum systems.  When a magnetic field is applied, each material presents some level of magnetization, and Quantum Mechanics says that magnetization is related to the expectation value of the spin angular momentum operator.  In the case of ferromagnetic materials, they can have a large magnetization even under the action of a small magnetic field and the magnetization process is always followed by hysteresis, and the magnetization is uniform and lined up with the magnetic field, usually these materials exhibit a strong ordering process that results in a parallel line up the spins\cite{Wang2015}.  In materials graphene type  we also can generate magnetic moment. In this form it is possible to study the transport phenomena \cite{McCreary}.
Overlapping between electronic wave functions are interactions well understood, again thanks to Quantum Mechanics, however there are other kinds of interactions occurring such as magnetocrystalline anisotropy, connected with the temperature dependence\cite{Zhuravlev} and demagnetization fields \cite{Flovik}, acting in low range.  
In such systems if we only consider the precession we will not reach the right limit. Certainly, the precession equation has to include a damping term providing the magnetization alignment with the magnetic field after a finite time \cite{Turek}.  In order to simulate these phenomena, several physical models have been presented. However, the Landau-Lifshitz model is still the one widely used in the description of the dynamics of ferromagnetic media.
In their pioneering work \cite{LL} in 1935, Landau and Lifshitz proposed a new theory based on the following dynamical equation: 
\begin{equation}
{\partial_t{\bf{M}}}= \vec{H}_{eff}\times \vec{M} +{{\alpha}\over{M^2_s}}\vec{M}\times (\vec{M}\times \vec{H}_{eff}), 
\end{equation}
where  $ \vec{H}_{eff}$ denote an effective magnetic field, with the gyromagnetic ratio absorbed, interacting with the magnetization $M = |\vec{M}|$.  The first term is the precession of the magnetization vector around the direction of the effective magnetic field and the second one describes a damping of the dynamics. With this theory we are able to compute the thickness of walls between magnetic domains, and also understand the domain formation in ferromagnetic materials. This theory, which now goes under the name of micromagnetics, has been instrumental in the understanding and development of magnetic memories. Landau and Lifshitz considered the Gibbs energy G of a magnetic material to be composed of three terms: exchange, anisotropy and Zeeman energies (due to the external magnetic field), and postulated that the observed magnetization per unit volume M field would correspond to a local minimum of the Gibbs energy. Later researchers added other terms to G such as magnetoelastic energy and demagnetization energy.  They also derived the Landau Lifshitz (LL) equation using only physical arguments and not using the calculus of variations.  In subsequent work,  Gilbert \cite{Gilbert} realized a more convincing form for the damping term, based on a variational approach, and the new combined form was then called  Landau Lifshitz Gilbert    (LLG) equation, today it is a fundamental dynamic system in applied magnetism. 

Nowadays, the scientific and technological advances provide a wide spectrum of manipulations to the spin degrees of freedom. The complete formulation for magnetization dynamics also include the excitation of magnons and their interaction with other degrees of freedom,  that remains as a challenge for modern theory of magnetism \cite{Iihama}.
These amazing and reliable kinds of procedures are propelling spintronics as a consolidated sub-area of Condensed Matter Physics \cite{Wolf}.   Since the experimental advances are increasingly providing high-precision data, many theoretical works are being presented \cite{Culcer,Mu,Yao,Mish} and  including strange materials, as the topological insulators, connected	 with the magnetocondutivity \cite{Adroguer} and graphene like structures \cite{Zhong} for a deeper understanding of the phenomenon including the spin polarization super currents for spintronics \cite{Eschrig:2015kya}, holographic understanding of spin transport phenomena \cite{Hashimoto:2013bna} and non-relativistic background \cite{Fadafan}.  

The torsion field  appears as one of the most natural extensions of General Relativity along with the metric tensor, which couples to the energy-momentum distribution,  inspects the details of the spin density tensor.  Actually, in General Relativity, fermions naturally couple to torsion by means of their spin.  

 In this work we consider that the torsion interacts with the matter in two types one of these is present in covariant derivative that contains the spin connection related with the Christoffell symbol given by the metric of the curve space time and  the contorsion given by the torsion that have  two 
antisymmetric index.  The other contribution is given by the non minimal spin torsion coupling that is important to the consistence of the theory.
It is possible to study the non relativistic approach to the torsion in connection with the spin particles  \cite{Bakke}, in this work we only consider the torsion contribution considering the plane space-time.

Our work is organized as follows, in Section 2,  we analyse the torsion coupling in relativistic limit,  in Section 3 the modified version of Pauli equation is presented.  Our approach starts with a field theoretical action where a Dirac fermion is non-minimally coupled in the presence of a torsion term, a low relativistic approximation is considered and the equivalent Pauli equation is then obtained. We derive a very similar expression to the Landau-Lifshitz-Gilbert (LLG) equation from our Pauli equation with torsion.  Under specific conditions for the magnetic moment, we show that the LLG equation can be established with damping  and dislocations terms.   

\section{The Relativistic and Non- Relativistic Discussions for Spin Coupling with Torsion}

In this Section, let us understand the way to describe the spin interaction by taking into account the torsion coupling. In our framework there are two terms in Dirac action, one of these is connected with spin current effect from the spin connection, the other is  a non-minimal spin torsion coupling whose effects are the subject of this work.   Dirac's equation is relativistic and we should justify why we use it in our model. Dispersion relations in Condensed Matter Physics (CMP) linear in the velocity appear in a wide class of models and one adopts the framework of Dirac's  equation to approach them. However, the speed of light, c, is suitably replaced by the Fermi velocity, $v_F$. Here, this is not what we are doing. We actually start off from the Dirac's equation and   we take, to match with effects of CMP, the non-relativistic regime, for the electron moves with velocities $ v\leq {c\over 300 }$. So, contrary to an analogue model where we describe the phenomena by a sort of relativity with c replaced by $v_F$, we here consider that the non-relativistic electrons of our system is a remnant of a more fundamental relativistic world. The non-relativistic limit is also more complete because it brings effects that do not directly appear in Galilean Physics. This is why we have taken the viewpoint of associating our physics to the Dirac's equation.

\subsection{The Dirac Model for Torsion and the Spin Current Interpretation}

 In this sub-section, we consider the microscopic discussion that gives the explicit form of the spin current in function of the gauge potential and torsion coupling. The scenario we are setting up is justified by the following chain of arguments: (i) We are interested in spin effects. We assume that there is a space-time structures (torsion) whose coupling with the matter spin becomes  relevant. But, we are actually interested in the possible non-relativistic effects stemming from this coupling, which is minimal and taken into account in the covariant derivative though the spin connection. (ii) The other point we consider is that, amongst the three irreducible torsion components, its pseudo-vector piece is the only one that couples to the charged leptons. Then, with this results in mind, we realize that  the electron spin density may non-minimally couple, in a Pauli- like interaction, to the field-strength of the torsion pseudo-vector degree of freedom. 
 
 So, our scenario is based on the relevant role space-time torsion, here modeled by a pseudo-vector, may place in the non-relativistic electrons of spin systems in CMP. The spin current that we talk  about is  the spin  magnetic  moment   and in general is not conserved alone. The quantity that is conserved is the  total magnetic moment that is the composition between both $\vec J = \vec{J}_S + \vec{J}_L$.  In form that $\partial_\mu J^\mu =0 $ where  the spin current  can be defined in related to the three component  spin current as  $J_S^\mu = \epsilon^{\mu a b}_{ \,\,\, \, \, \, \, \, \, \rho} \,  J^{\rho}_{\, \, \, a b }$ and $\vec{J}_L$ is the spin orbit coupling .  In analogy with the charge current, defined by the derivative of the action in relation to the gauge field $A_\mu$  we used the definition where the spin current is  the derivative of the action in relation with the spin connection. We consider here the spin current as the derivative of the action in relation to the spin connection $\omega_{\mu}^{\, \, \, a b}$ then the spin current is given by
\begin{equation}
J^\mu_{\, \, \, a b} = {\delta S \over \delta \omega_{\mu}^{\, \, \, a b}} ,
\end{equation}
where $ \omega_{\mu}^{\, \, \, a b } $ is 
\begin{equation}
\omega_{\mu}^{\, \, \, a b} = e^{\,\,a}_\nu \nabla_\mu e^{\nu b} + e_\lambda^{\, \, a} \Gamma^{\, \, \, \, \, \,  \lambda}_{\mu \nu} e^{\nu b}  + e_\lambda^{\,\,a }\, K^{\, \, \, \,\, \lambda}_{\mu \nu} \, e^{\nu b},
\end{equation}
and $   K^{\, \, \, \, \lambda}_{\mu \nu}   $ is the contorsion  given by
\begin{equation}
K^{\, \, \, \,   \, \lambda}_{\mu \nu}  = - {1\over 2} (T_{\mu \, \, \, \nu}^{\, \, \lambda} + T_{\nu \, \, \, \mu}^{\, \, \lambda} - T_{\mu  \nu}^{\, \, \, \, \,  \lambda}),
\end{equation}
where $T_{\mu  \nu}^{\, \, \, \, \,  \lambda}$ is the torsion. 
In this work we consider two terms for torsion, on of these  is the totally anti symmetric tensor, that respecting the duality relation given by $
T_{\mu  \nu \lambda} = \epsilon _{\mu  \nu  \lambda \rho}  \, S^\rho $ where $S_\rho $ is the pseudo-vector  part of torsion.\footnote{Considering space times with torsion $T^{\alpha}_{\beta \gamma}$, the afine connection is not symmetric, $T^{\alpha}_{\beta \gamma}=\Gamma{^{\alpha}_{\beta \gamma}-\Gamma{^{\alpha}_{\gamma \beta}}}$, and we can split it into three irreducible components, where one of them is the pseudo-trace $S^{\kappa}={1\over6}\epsilon^{\alpha \beta \gamma \kappa}T_{\alpha \beta \gamma}$.} 
The other term that we consider is the 2-form tensor $ T^{\mu \nu}$ where $T_{\mu \nu} = \partial_\mu S_{\nu } - \partial_\nu S_{\mu} 	$ this term is analog to the field strength of the electromagnetic gauge potential $A_{\mu}$  that in our case is changed to pseudo-vector $S_{\mu} $. The invariant  fermionic action that contained these contributions for torsion  is given by
\begin{eqnarray}
S=\int d^4x i\bar{\psi}  ( \gamma^{\mu}{\cal D}_{\mu} +   \lambda T_{\mu \nu} \Sigma^{\mu \nu}  +\,m  ){\psi}   ,\label{dirac}
\end{eqnarray}
where the covariant derivative is ${\cal D}_{\mu}= D_\mu - i \eta \, \omega_\mu^{\, \, \, \, a b}  \Sigma_{ab}   $ that contain the covariant gauge derivative $ D_\mu  = \partial_{\mu}-ieA_{\mu}   $ and the spin connection covariant derivative. 

 We consider the flat space-time where the only contribution for the spin connection is the contortion. In this form we have a spin current given by
\begin{equation}
J^\mu_{\, \, \, a b} = {1 \over 2} \bar \psi  \gamma^\mu \Sigma_{ab} \psi .
\end{equation}
 We consider the ansatz where $ \omega_\mu^{\,\,\, a b} $ contain the total  antisymmetric part of the contorsion given by
 \begin{equation}
\omega_\mu^{\,\,\, \kappa \lambda}  = \mu \, \epsilon_{\mu}^{ \,\, \,\kappa \lambda \rho} S_\rho  .\end{equation}
 We used the splitting $
 \gamma^{\mu } \Sigma _{\kappa \lambda}   =  \epsilon^\mu_{\,\, \kappa \lambda\rho }\gamma^\rho \gamma_5  + \delta^{\mu}_{\kappa} \gamma_{\lambda} - \delta^{\mu}_{\lambda} \gamma_{\kappa} $
 That give us the current in the form 
 \begin{equation}
J^\mu_{\, \, \, \alpha \lambda} = {\delta S \over \delta \Gamma_\mu^{\,\,\, a b}} = {1 \over 2}  \epsilon^\mu_{\,\, \alpha \lambda \rho } \bar \psi  \gamma^\rho \gamma_5 \psi =  J^i_{\, \, \, j k}    +  J^0_{\, \, \, i j} 
\end{equation}
 where the currents are
 \begin{eqnarray}
J^i_{\, \, \, j k} &=& {1\over 2}  \epsilon^i_{\,\, j k  } \bar \psi  \gamma^0 \gamma_5      \psi  ;\\
J^0_{\, \, \, i j} &=& {1\over 2}   \epsilon_{\,\, i j k  } \bar \psi  \gamma^k \gamma_5  \psi ,
\end{eqnarray}
then the current part of the action coming from the covariant derivative is given by

\begin{equation}
S_{curr}  =  - \eta \int  d^4x  \gamma^{\mu} \gamma_5    S_\mu = \eta \int d^4x \, \, \vec{S} \cdot \vec{J}.
\end{equation}

The action (\ref{dirac}) considers  the temporal component of the torsion pseudo-vector $S_0 = 0 $, we have can be written as
\begin{eqnarray}
S  =\int \,\,d^4x i\bar{\psi}  \Big( \,\gamma^{\mu} \partial_\mu \!\!+ ig  \gamma^{\mu}\! A_\mu  \!\! + i\eta \gamma^{\mu} \gamma_5    S_\mu \!\!\!   + \!\!  \lambda T_{\mu \nu} \Sigma^{\mu \nu} \!\!\! +\!m  \,\Big){\psi}   ,\label{dirac1}
\end{eqnarray}
Our sort of gravity background does not exhibit metric fluctuations. The space-time is taken to be flat, and we propose a scenario such that the type of gravitational background is parametrized by the torsion pseudo-trace $S_{\mu}$, whose origin may be traced back to one geometrical defect.

\subsection{Dirac equation in presence of torsion}

Now, we discuss the Dirac equation given by eq. (\ref{dirac1}). From the action above, taking the variation with respect to $(\delta S/\delta\bar{\psi})$, a modified Dirac's equation reads as below:
\begin{eqnarray}
\label{dirac equation}
\big[i\gamma^{\mu}\partial_{\mu}-\eta\gamma^{\mu}\gamma_{5}S_{\mu}-e\gamma^{\mu}A_{\mu} + \lambda \Sigma^{\mu \nu} \partial_\mu S_\nu +m\big]\psi=0.  \label{Dirac}  
\end{eqnarray}
For a vanishing $\lambda-$parameter,  the equation (\ref{dirac equation}) has been carefully studied in \cite{Sha1,Sha1a,Hamm1,Sha2,Sha4}.

The generation, manipulation, and detection of a spin current, as well as the flow of electron spins, are
the main challenges in the field of spintronics, which involves the study of active control and
manipulation of the spin degree of freedom in solid-state systems, \cite{Zutic,Wolf,Maekawa}. A spin current
interacts with magnetization by exchanging the spin-angular momentum, enabling the direct
manipulation of magnetization without using magnetic fields \cite{Grollier, Ando}. The interaction between spin currents and magnetization provides also a method for spin current
generation from magnetization precession, which is the spin pumping \cite{Tserkovnyak, Mizukami}. We showed in last Sections that there are two type of deformations induced by the torsion. Both of these can be generating a spin current.  

After a suitable separation of components $(\mu=0,1,2,3)$, the equation of motion can be written as,
\begin{eqnarray}
i{\partial_t \psi} & = &i\alpha_i{\partial_i \psi}+\eta \gamma_5 S_0 \psi-\eta \alpha_i \gamma_5 S_i\psi+eA_0\psi \, +\nonumber \\
&-&  e\alpha_i A_i \psi-{i \lambda\over4}\epsilon_{ijk}\beta \gamma_5 \alpha_k \partial_i S_j \psi +\beta m \psi  .
\end{eqnarray}

Defining a gauge-invariant momentum, $\pi_j=i\partial_j-eA_j$, and using that $\Sigma_{ij}=-{i\over4}\epsilon_{ijk}\gamma_5 \alpha_k$ with $\gamma_0 = \beta$, the effective Hamiltonian takes the form,
\begin{eqnarray}
\label{Hamiltonian}
H&=&\alpha_k\pi_k + \eta \gamma_5 S_0 -\eta \alpha_k \gamma_5 S_k + e A_0  \, + \nonumber \\
&- & {{i\lambda}\over{4}} \epsilon_{ijk}\beta \gamma_5 \alpha_k \partial_i S_j+\beta m\,.
\end{eqnarray}
where we have  the matricial definitions:

\begin{eqnarray}\alpha_i&=&\left(
\begin{array}{cc}
0 & \sigma_i \\
\sigma_i & 0
\end{array}\right), \gamma_5\,=\,\left(
\begin{array}{cc}
0 & 1 \\
1 & 0
\end{array}\right), \beta\,=\,\left(
\begin{array}{cc}
1 & 0 \\
0 & -1
\end{array}\right),
\end{eqnarray}

In the Heisenberg picture, the position, $\vec{x}$, and momentum, $\vec{\pi}$, operators obey two different kinds of relations; we consider the torsion as a function of position only, $S=S(\vec{x})$, so that
\begin{eqnarray}
\dot{\vec{x}}&=&\vec{\alpha} \nonumber \\
\dot{\vec{\pi}}&=&e(\vec{\alpha} \times \vec{B})+e \vec{E}+\eta \gamma_5
{{\partial}\over{\partial x}}(\vec{\alpha} \cdot \vec{S})\hat{x}
\,.
\end{eqnarray} 
One reproduces the usual relation for $\dot{\vec{x}}$, while the equation for $\dot{\vec{\pi}}$ presents a new term apparently giving some tiny correction to the Lorentz force.  

However, if we consider the torsion in a broader context, now as a momentum- and position-dependent background field, $S=S(\vec{x},\vec{k})$, we have to deal with the following picture,
\begin{eqnarray}
\dot{\vec{x}}&=&\vec{\alpha}-\eta \gamma_5
{{\partial }\over{\partial k}}(\vec{\alpha} \cdot \vec{S})\hat{k} \nonumber \\
\dot{\vec{\pi}}&=&e(\vec{\alpha} \times \vec{B})+e \vec{E}+\eta \gamma_5
{{\partial}\over{\partial x}}(\vec{\alpha} \cdot \vec{S})\hat{x} \, + \nonumber \\ &+&  e \eta \gamma_5  {{\partial A}\over{\partial x}}{{\partial}\over{\partial k}}(\vec{\alpha} \cdot \vec{S})\hat{k}
\,. 
\end{eqnarray}

The two sets of dynamical equations above are clearly showing us the small corrections induced by the torsion term.  Nevertheless, we are still here in the  relativistic domain, and it is necessary change this framework for a better understanding of the SHE phenomenology.  For this reason, in next Section we are going to approach the system by going over into its non-relativistic regimen

\subsection{Non-Relativistic Approach with torsion}

In this sub-section, we consider the Dirac equation in its non-relativistic limit.  
One important requirement for the Dirac equation is that it reproduces what we know from non-relativistic quantum mechanics.  We can show that, in the non-relativistic limit, two components of the Dirac spinor are large and two are quite small. To make contact with the  non-relativistic description , we go back to the equations written in terms of $\varphi$ and $\chi$ of the four component spinor $\psi = {e^{i\,m\, t} \over \sqrt{2m}}\left(\begin{array}{ll}\varphi \\
\chi
\end{array}\right) $, just prior to the introduction of the $\gamma $ matrices. we obtain two equation on of these for $\varphi$ and the other for $\chi$. We can solved in $\chi $ and substiuted in the Dirac equation given by (\ref{dirac1}) and take the 
 non-relativistic regimen $(\mid \vec{p}\mid << m)$.
So, in this physical landscape, from now and hereafter, our goal is to consider a low-relativistic approximation based on an extended Pauli equation version by including torsion as presented before.  Employing the Hamiltonian (\ref{Hamiltonian}), we could carry out our calculations in the framework of the Fouldy-Wouthuysen transformations; however for the sake of our approximation at lowest order in $v/c$, we take that SHE is adequately well described by the low-relativistic Pauli equation.  We are considering that the electron velocities are in the range of Fermi's velocity.   In this case, we  arrive at the version given below for the Pauli's equation:    
\begin{eqnarray}
\label{pauli0}
i{{\partial \varphi}\over{\partial t}}&=& \Big[{{(\vec{p}-e\vec{A})^2}\over{2m}}-{{e}\over{2m}}(\vec{\sigma}_{eff} \cdot \vec{B})+eA_0 - ({\bf \sigma} \cdot \vec{S}_{eff})+\nonumber \\
&-& {{\lambda}\over{8m}}(\vec{\nabla}\times {\vec{S}})\cdot(\vec{\sigma}\times \vec{p})+ {{i\lambda}\over{8m}}(\vec{\nabla}\times {\vec{S}})\cdot \vec{p}+\nonumber \\
&-&{{e\lambda}\over{8m}}(\vec{\nabla}\times {\vec{S}})\cdot(\vec{\sigma}\times {\vec{A}}) 
+{{ie\lambda}\over{8m}}(\vec{\nabla}\times {\vec{S}})\cdot \vec{A}\Big]\varphi\,. 
\end{eqnarray} 

The equation above displays the usual Pauli terms, but corrected by new terms due to the torsion coupling.  The second and the fourth contributions in the RHS of eq.(\ref{pauli0}) can be thought of as effective terms for $\vec{\sigma}$ and $\vec{S}$, respectively given by    
\begin{eqnarray}
{\bf{\sigma}}_{eff}& =& \vec{\sigma} +{\eta \over 2 m}\vec{S} +i {\eta \over 2m} (\vec{\sigma}\times \vec{S} )\\
\vec{S}_{eff}& =& \eta \vec{S} +i{\lambda \over 4}(\vec{\nabla}\times \vec{S} ).
\end{eqnarray}
The fifth contribution in the RHS is proportional to the Rashba SO coupling term; this term yields an important effect on the behavior of spin.  

\section{From the Modified Pauli Equation to Unfold in LLG}

In this Section, we consider the magnetization equation derivation given by Dirac non-relativistic limit take into  account the presence of torsion. Let us start by considering the modified Pauli equation eq.(\ref{pauli0})  and find the magnetization equation. By using the Landau gauge $\vec{A} = H \, \vec{x}$ and taking that $\vec{x}\cdot \vec{\sigma} =0$ (the spins are aligned orthogonally to the plane of motion), give us the Hamiltonian of the full system in the non-relativistic limit as:
\newpage
\begin{eqnarray}
\label{pauli1a}
H&\!=\!&{{(\vec{p}-e\vec{A})^2}\over{2m}}-{{e}\over{2m}}(\vec{\sigma}_{eff} \cdot \vec{B})+eA_0 - (\sigma \cdot \vec{S}_{eff})
- {{\lambda}\over{8m}}(\vec{\nabla}\times {\vec{S}})\cdot(\vec{\sigma}\times \vec{p})+ {{i\lambda}\over{8m}}(\vec{\nabla}\times {\vec{S}})\cdot \vec{p}, \nonumber \\
\end{eqnarray} 
with $\vec{B}(t) = \mu_0 \vec{H}(t)$, where $\mu_0$ is the gyromagnetic ratio.   
 The Pauli equation associated with (\ref{pauli1a}) reads as follows below

\begin{eqnarray}
\label{pauli1}
i{{\partial \varphi}\over{\partial t}}&=& \Big[{(\vec p - e A  )^2 \over{2m}}-{{e}\over{2m}}(\vec{\sigma}_{eff} \cdot \vec{H})+eA_0 - (\vec{\sigma} \cdot \vec{S}_{eff})  + \nonumber \\
&- & {\lambda \over 8m} (\vec\nabla \times \vec{S})\cdot  (\vec\sigma \times \vec p))  + {i \lambda \over 8m}(\vec \nabla \times \vec S) \cdot p \Big] \varphi .
\end{eqnarray}

 Let us consider the magnetization vector equation related win the spin magnetic moment $\vec{\mu} = {e \over 2m} \vec{\sigma} $. 
In our approach we consider the magnetization is defined by $\vec{M}= (\vec{\mu} \, \varphi)^\dagger \, \varphi -\varphi^\dagger \, (\vec{\mu} \, \varphi) $ where $\varphi $ given by Pauli equation (\ref{pauli1}) and $\varphi^\dagger \varphi =1$ and $\vec{\hat S} \varphi = \vec{S} \varphi $, with the notation $\vec{\hat S} $ is a torsion operator and $ \vec{S} $ is the torsion autovalue.
 We have, by the manipulation of Pauli equation   the magnetization equation associated with a fermionic state when we applied a external magnetic field $\vec{H}$ considering the Pauli product algebra as ${1 \over 2}( \sigma_i \, \sigma_j - \sigma_j \sigma_i) = i \epsilon_{i j k} \sigma_k $ and 
${1 \over 2}( \sigma_i \, \sigma_j + \sigma_j \sigma_i)  = \delta_{i j} $. The magnetization equation that arrive that is 

   \begin{eqnarray}
{\partial \vec{M} \over \partial t} &=  &\vec{M}  \times \vec{H}   + \eta  \vec{M}  \times \vec{S}  +\beta (\vec{M} \times \vec{L}) \, + \nonumber\\
& +&  \eta{e\over 2 m^2} (\vec{S} \times \vec{H})  \, +{ e \lambda \over 2 m} (\vec \nabla \times \vec S ),\label{mag} 
\end{eqnarray} 
 with the magnetic moment given by $\vec{L} = \vec{r} \times \vec{p}$ and  $\vec{S} $  as the torsion pseudo-vector.   We can observed that there are two terms that arrived  by the covariant derivative ${\cal D}_\mu$ defined by the coupling constant $\eta $ and the other is the parameter that arrived by non-minimal spin torsion coupling with the coupling constant $\lambda$.  
Where the effect of the new terms given when $\eta \neq 0 $ and $\lambda \neq 0$.  
 
 We consider the scalar product of the magnetization $\vec{M}$, the magnetic field $\vec{H}$ and the torsion pseudo-vector $\vec{S}$ with the equation (\ref{mag}) and we obtain\footnote{We used the vectorial relating given by $A\cdot (B \times C)= B \cdot (C \times A) = C\cdot (A \times B)$ the other is  $A \times (B \times C) =  (A\cdot C) B - (A \cdot B) C$  and $ \nabla \cdot ( A \times B) = B \cdot (\nabla \times A) - A \cdot (\nabla \times B)$. }

\begin{equation} 
\partial_t\Big[{1 \over 2}(\vec{M}\cdot \vec{ M}) \Big] = {\eta e \over 2m^2} \Big[  \vec{M} \cdot (\vec{S} \times \vec{H} ) +{\lambda m \over \eta}\vec{M} \cdot (\vec{\nabla}\times \vec{S})\Big]; \label{damping1}
\end{equation}
\begin{eqnarray} 
\partial_t\Big[{1 \over 2}(\vec{H}\cdot \vec{ M}) \Big] &=&  \Big[\eta\,  \vec{M} \cdot (\vec{S} \times \vec{H} ) + \beta \vec{M} \cdot (\vec{L} \times \vec{H}) \, +\nonumber\\&+& {e \over 2 m }\lambda\vec{H} \cdot (\vec{\nabla}\times \vec{S})\Big]; \label{damping2}
\end{eqnarray}
\begin{equation} 
\partial_t\Big[{1 \over 2}(\vec{S}\cdot \vec{ M}) \Big] = \vec{M}\cdot (\vec{S} \times \vec{H} )  +\beta \vec{M}\cdot (\vec{L} \times \vec{H} ).\label{damping3}
\end{equation}

  \begin{eqnarray}
\partial_t\Big[{1 \over 2}(\vec{L}\cdot \vec{ M}) \Big]&=  &\vec{M} \cdot (\vec{L}  \times \vec{H} )  + \eta  \vec{M} \cdot (   \vec{L} \times \vec{S}) \, + \nonumber\\
& +&  \eta{e\over 2 m^2} \vec{S} \cdot (\vec{L} \times \vec{H})  +{ e \lambda \over 2 m} \vec{L} \cdot (\vec \nabla \times \vec{ S} ),\label{mag} .
\end{eqnarray} 
 With the equations dysplayed in (\ref{damping1})-(\ref{damping3}), it is possible to inspect the general behavior of the magnitude of the magnetization, $\vec{M} $, that precesses around the magnetic field, $ \vec{H} $. In our framework, the magnetization also  precesses around the  torsion vector $\vec{M} \cdot  \vec{S} $.  
   Without torsion, we have  ${\partial \vec{M} \over \partial t} =  \, \vec{M}  \times \vec{H} $, so that $\vec{M} \cdot \vec{M}$= constant and  $\vec{M}\cdot \vec{H} $= constant as in the usual case of the electron under the action of a time-dependent external magnetic field, with the Zeeman term given by the Hamiltonian $H_M =  \vec{M} \cdot \vec{H}$. 
    
\subsection{Planar torsion analysis with damping}

Here, we intend to analyze some possibilities of solutions to the magnetization that respect the conditions given by (\ref{damping1})-( \ref{damping3}). 
The magnitude of the magnetization is not constant in general, as we can see in equation (\ref{damping1}), but, if this quantity is constant, there comes out a constraint given by
 
 \begin{equation}
 \vec{M} \cdot (\vec{S} \times \vec{H}) = -{\lambda \, m \over \eta} \vec{M} \cdot (\vec{\nabla} \times \vec{S}).
 \end{equation}
 If we consider ${d(\vec{L}\cdot \vec{M} )\over dt}=0$, \, \,  we have 
 \begin{eqnarray}
 \vec{S} \cdot (\vec{L} \times \vec{H})  = -{ m \lambda \over \eta} \vec{L} \cdot (\vec \nabla \times \vec{ S} ).
  \end{eqnarray}
This expression describes us the case where $\vec{M} \cdot \vec{H} \neq 0$ and $\vec{S} \cdot \vec{M} \neq 0$;  then, there is a the damping  angle in both directions given by the precession around the magnetic field $\vec{H}$ and around the torsion pseudo-vector $\vec{S}$ :
\begin{equation} 
\partial_t\Big[{1 \over 2}(\vec{H} \cdot \vec{ M}) \Big] = \lambda  m({e \over 2m^2}\vec{H}-\vec{M})  \, \cdot (\vec{\nabla}\times \vec{S}) ; \label{damping4}
\end{equation}
\begin{equation} 
\partial_t\Big[{1 \over 2}(\vec{S}\cdot \vec{ M}) \Big] =  - {\lambda \, m \over \eta}  \, \vec{M} \, \cdot (\vec{\nabla}\times \vec{S}).\label{damping5}
\end{equation}

Let us  consider the first proposal in a very particular and very simple case for a planar torsion field, $\vec{S}={1\over2}\chi(x\hat{y}-y\hat{x})$; this choice allows us to realize the curl of torsion as an effective magnetic field, $\vec{\nabla}\times {\vec{S}}=\vec{B}_{eff}=\chi\hat{z}$. 

If we pick up the configuration of Fig. 1, we  find the relation of the angle between the magnetic field $\vec{H} $ and the magnetization $\vec{M} $.  

 \begin{figure}[htb]
\begin{center}
\includegraphics[width=8cm, height=8cm]{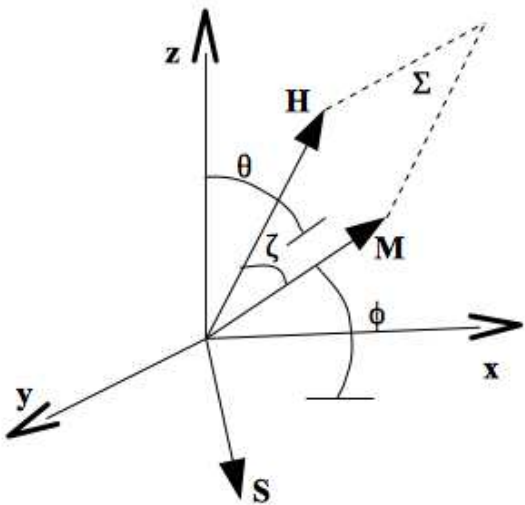}
\caption{ Magnetization vector rotating around the magnetic field $\vec{H} $ with damping  given by the dynamics of the angle $\zeta $.  The system $\{M, H\}$ rotates around the vector $\vec{S}$ in the xy-plane also with damping given by the angle $\phi$.  We consider $\phi =  \omega_{\phi} t$ with $\omega_\phi= \omega_\theta = {2 \lambda m \chi \over \eta S}$; with this configuration $\zeta = \omega_\zeta t = {\lambda \chi m\over \eta HM} \Big(H + \eta M \Big)t $.
 }
\end{center}
 \label{Figure1}
 \end{figure}

 We started off by discussing the case where the torsion is planar with the magnitude of the magnetization being constant, $ \vec{M} \cdot \vec{M} =0$, and
 
\begin{equation} 
\partial_t\Big[{1 \over 2}(\vec{S}\cdot \vec{ M}) \Big] =  - {\lambda \chi \, m \over \eta} \vec{M}  \cdot \vec{z},\label{damping6}.
\end{equation}
this gives us  the  magnetic momentum precession around the torsion. For consistency, we show that this result is compatible with the equation 
\begin{equation} 
\partial_t\Big[{1 \over 2}(\vec{H} \cdot \vec{ M}) \Big] = \lambda  m  \chi ({e \over 2m^2}\vec{H}-\vec{M})   \, \cdot  \vec{z}. \label{damping7}
\end{equation}
 In the case of the Fig. 1, the magnetization precesses around the magnetic field and around the planar torsion vector both with damping.

\subsection{Helix-Damping Sharped Effect in a Planar Torsion Configuration }

Now, let us consider the most general case, where the magnitude of the magnetization is not constant, but with $(\vec{L}\cdot \vec{M} )=0$. The configuration is considered in  Fig. 2. 
\begin{figure}[htb]
\begin{center}
\includegraphics[width=10cm, height=8cm]{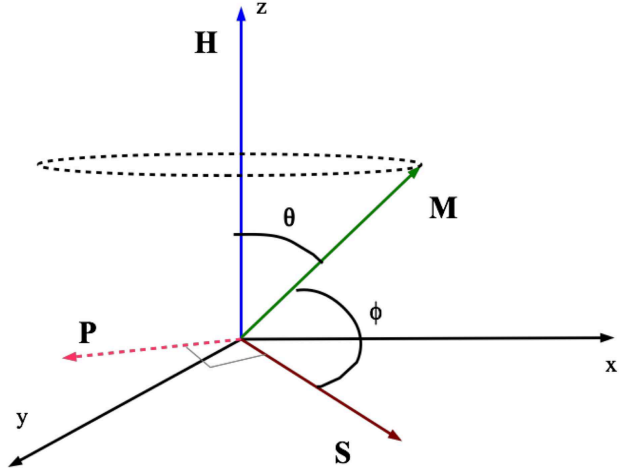}
\caption{In this picture we show the effect of the torsion in magnetization dynamics. The green vector is the magnetization vector and the blue vector is the external magnetic field. }
\end{center}
 \label{Figure2}
 \end{figure}
 With the expressions (\ref{damping1})-(\ref{damping3}), we can readily write the magnitude of the magnetization 
  
\begin{equation} 
\partial_t\Big[{1 \over 2}(\vec{M}\cdot \vec{ M}) \Big] = {e \eta \over 2m^2} \Big[  \partial_t\Big[{1 \over 2}(\vec{S}\cdot \vec{ M}) \Big]  +{\lambda m \over \eta}\vec{M} \cdot (\vec{\nabla}\times \vec{S})\Big]. \label{damping4}
\end{equation}
 By using of the equation (\ref{damping4}) and considering $ \partial_t\Big[{1 \over 2}(\vec{S}\cdot \vec{ M}) \Big] \neq 0$, we can see, the example of the Fig. 3, that $\partial_t\Big[{1 \over 2}(\vec{M}\cdot \vec{ M}) \Big] \neq 0$. This possibility gives us that the magnitude of magnetization is not constant, as in the usual LLG. This effect is the effect of torsion that gives us that the rotational lines do not return around themselves.  
\begin{figure}[htb]
\begin{center}
\includegraphics[width=10cm, height=8cm]{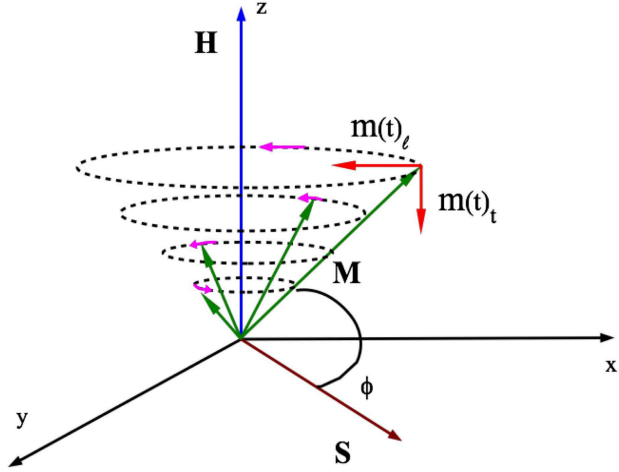}
\caption{In this draw we show the effect of the torsion in magnetization dynamics. The green vector is the magnetization vector and the blue vector is the external magnetic field. In this representation we used $|\vec{M}| = M(t)$, $|\vec{S}| =$ constant  and $|\vec{H}| =$ constant with $\omega_\theta = {2 m \lambda \chi \over \eta S}$  then  $M(t) = {\lambda e \chi \over \omega_\theta m} \sin \omega_\theta t $. }
\end{center}
 \label{Figure3}
 \end{figure}
 
  \begin{figure}[htb]
\begin{center}
\includegraphics[width=6cm, height=6cm]{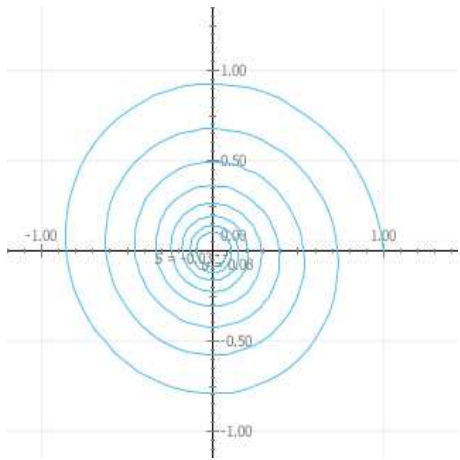}
\caption{ Damping behavior in torsion plane. Show the behavior of the $\phi =\omega_\phi t$ dynamic. }
\end{center}
 \label{Figure4}
 \end{figure} 

 Equation (\ref{damping4} ) does not involve the explicit dependence of the magnetic field. We choose to work out the equation

\begin{equation} 
\partial_t\Big[{1 \over 2}(\vec{H}_{eff}\cdot \vec{ M}) \Big] =   {e \over 2 m }\lambda \chi \vec{H} \cdot  \vec{z}, \label{damping5}
\end{equation}
where $\vec{H}_{eff} = \vec{H} - \eta \vec{S}$ gives us the explicit form of the magnetic field interaction with the magnetization. 

We notice that this quantity is different from zero, then the angle between the magnetic field and the magnetization is not constant; this yields us the damping precessing effect of the magnetization vector around the magnetic field. The composition between these two effects, dislocation and damping, is what we refer to as the  helix-sharped with damping, effect where the damping effect  can be see in Fig. 4.
We can show that there are two magnetization effects: the damping given by the longitudinal magnetization function  $m(t)_{l}$ and dislocations given by the longitudinal magnetization function $m(t)_t$ as we can see in Fig. 2.
The trajectory of the magnetization is the conical increasing spiral,  where the modulus of magnetization increases with the time. In the torsion plan the behavior is given by Fig3.

\newpage

\section{Concluding Remarks}
 
 In this work, we have considered  that the magnetization equation is a non-relativistic remnant of the  non-relativistic limit of the Dirac equation with torsion couplings. We have considered two types of couplings: one of these related with the spin current in Dirac equation, defined by the spin connection. When we derived the action in relation with the spin connection we obtain the spin current, this description is analog to the charged current when we have the derivation of the action in relation with the gauge field. 
 
We refer to the other term as the non-minimal torsion term and It gives us the rotational of the torsion. We have analyzed this term in the general context and observed that it is possible to recover the Landau Lifshitz in the case were the torsion is zero. Then, we can point out that the non-relativistic limit of the Dirac equation reproduces the usual case where the magnetization vector precesses around the magnetic field. When we introduce the torsion terms we analyze, in the general regime the magnitude of magnetization $\vec{M} \cdot \vec{M} $, the precession of the magnetization around the magnetic field $\vec{H}\cdot \vec{M}$, and the precession of the magnetization around the torsion pseudo-vector $\vec{S}\cdot \vec{M} $ is not constant. 

When the magnitude of the magnetization is constant, in the case where the torsion is planar, there are two possible magnetization precessions one around the magnetic field and other around the planar torsion pseudo-vector. In both dynamics, there occurs damping. An interesting example has been analyzed in Fig. 1, where we show that it is possible to realized an apparatus in some experimental device. In this sense, our framework can reproduce the LLG equation.  
 The most general approach should consider that the magnitude of the magnetization is not constant.  In this case, as we can see from Fig. 2, the loop drawn by the magnetization damping but the is not remain in the same plane. This effect is typically a torsion effect, were the lines are not closed. This effect seems to be like a dislocation in the material that presents topological defects like solitons and vortices.   Both dislocation  and damping give us what  we  refer to as the helix-damping sharp effect, wich is a new feature of the models with torsion\cite{S.:2001sm}.   
  
We can observe that this result is the new feature introduced by the planar torsion, if we consider the comparation with damping and dislocations  terms presented in  LLG equation. This may help in the task of setting up new apparatuses and maybe experimental purposes to explore such characteristics in this phenomenon.  We have  found that $\vec{M} \cdot \vec{M}$
= constant,  its consequence  is the dislocation effect. The damping effect is the usual one, where  the angle dynamic can crease and decrease with the time.  In this work we does not study the polarization of the spins that is subject of next work when we will consider these systems in terms or the spin up and spin down dynamic. In the literature, this effect is named pumped spin current \cite{Andreev}, and we shall study the possibility of this current when the system is in a helix-sharped configuration\cite{Yoda}.


\begin{thebibliography} {99}



\bibitem{Dyrdal}
A. Dyrdal, J. Barnas, Phys. Rev. B {\bf 92}, 165404 (2015)

\bibitem{Wang2015} Z. Wang, C. Tang, R. Sachs, Y. Barlas, and J. Shi, Phys. Rev. Lett. {\bf 114}, 016603 (2015).

\bibitem{McCreary} K. M. McCreary, A. G. Swartz, W. Han, J. Fabian, R.K. Kawakami Phys. Rev. Lett. {\bf 109}, 186604 (2012)


\bibitem{Zhuravlev} I. A. Zhuravlev, V. P. Antropov and K. D. Belashchenko, Phys. Rev. Lett. {\bf 115}, 217201 (2015).

\bibitem{Flovik} V. Flovik, F.  Maci�, J. M. Hern�ndez, R. Brucas, M. Hanson and E. Wahlstr�m, Phys. Rev. B {\bf 92}, 104406 (2015).

\bibitem{Turek} I. Turek, J. Kudrnovsky and V. Drchal, Physical Review B {\bf 92}, 214407 (2015).

\bibitem{LL} L.D. Landau, E.M. Lifshitz, "On the theory of the dispersion of magnetic permeability in
ferromagnetic bodies", Phys. Z. Soviet Union {\bf 8}, 153 (1935); M. Lakshmanan, "The fascinating world of the Landau Lifshitz Gilbert equation: an overview", Phil. Trans. R. Soc. A {\bf 369}, 1280  (2011). 



\bibitem{Gilbert} Gilbert, T. L.  "A phenomenological theory of damping in ferromagnetic materials", IEEE Trans. Magn. {\bf 40}, 34433449 (2004).

\bibitem{Iihama} S. Iihama, S. Mizukami, H. Naganuma, M. Oogane, Y. Ando, and T. Miyazaki, Phys. Rev. B {\bf 89}, 174416 (2014).

\bibitem{Wolf} S. A. Wolf, D. D. Awschalom, R. A. Buhrman, J. M. Daughton,S. von Moln\'ar,M. L. Roukes, A. Y. Chtchelkanova,D. M. Treger , Science  {\bf 294} ,1488  (2001).

\bibitem{Culcer}D. Culcer et al., Phys. Rev. Lett. {\bf 93},  046602 (2004).

\bibitem{Yao} J. Yao and Z. Q. Yang, Phys. Rev. B  {\bf 73}, 033314  (2006).
 
\bibitem{Mish} E. G. Mishchenko, A.V. Shytov and B. I. Halperin Phys. Rev. Lett. {\bf  93}, 226602 (2004).

\bibitem{Mu} Sh. Murakami, N. Nagaosa, S.-C. Zhang, Science  {\bf 301}, 1348 (2003).


\bibitem{Adroguer} P. Adroguer, E. L. Weizhe,  D. Culcer and E. M. Hankiewicz, Phys. Rev. B {\bf 92}, 241402 (2015).

\bibitem{Zhong} Xin-Zhong Yan and C. S. Ting, Physical Review B {\bf 92}, 165404 (2015) .

\bibitem{Eschrig:2015kya} 
  M.~Eschrig,
  Rept.\ Prog.\ Phys.\  {\bf 78}, no. 10, 104501 (2015).

\bibitem{Hashimoto:2013bna} 
  K.~Hashimoto, N.~Iizuka and T.~Kimura,
  Phys.\ Rev.\ D {\bf 91}, no. 8, 086003 (2015).

\bibitem{Fadafan} K. B. Fadafan and F. Saiedi, Eur. Phys. J. C  {\bf 75} 612 (2015).

\bibitem{Bakke} K. Bakke, C. Furtado and J. R. Nascimento, Eur. Phys. J. C {\bf 60}, 501 (2009) : erratum Eur. Phys. J. C {\bf 64} 169 (2009) 

\bibitem{Sha1} I.L. Buchbinder and I.L. Shapiro, Phys. Lett. B {\bf151} 263 (1985)   

\bibitem{Sha1a}V.G. Bagrov, I.L. Buchbinder and I.L. Shapiro, Sov. J. Phys. {\bf 35} 3 (1992), hep-th/9406122

\bibitem{Hamm1} R. T. Hammond, Phys. Rev. D {\bf52} 12 (1995).

\bibitem{Sha2} I.L. Shapiro, Phys. Rep. {\bf357} 113 (2002) 


\bibitem{Hamm} R. T. Hammond Rep. Prog. Phys. {\bf65} 599 (2002). 

\bibitem{Sha4} L.H. Ryder and I.L. Shapiro, Phys. Lett. A  {\bf247} 21 (1998)


\bibitem{Zutic}  Zutic I, Fabian J and Sarma S D,  Rev. Mod. Phys. {\bf 76}, 323 (2004).


\bibitem{Maekawa} Maekawa S., Concepts in Spin Electronics (Oxford: Oxford University Press), (2006). 


\bibitem{Grollier}  Grollier J, Cros V, Hamzic A, George J M, Jaffr`es H, Fert A, Faini G, Youssef J B and Legall H., Appl.Phys. Lett. {\bf 78}, 3663 (2001). 

\bibitem{Ando}  Ando K, Takahashi S, Harii K, Sasage K, Ieda J, Maekawa S. and Saitoh E., Phys. Rev. Lett. {\bf 101}, 036601 (2008). 

\bibitem{Tserkovnyak}Tserkovnyak Y, Brataas A. and Bauer G. E. W. , Phys. Rev. Lett. {\bf 88}, 117601 (2002).



\bibitem{Mizukami} Mizukami S, Ando Y and Miyazaki T., Phys. Rev. B {\bf 66} 104413 (2002).


















\bibitem{S.:2001sm} 
  S.~Azevedo,
  J.\ Phys.\ A {\bf 34}, 6081 (2001).


\bibitem{Andreev} P. A. Andreev, Phys. Rev. E {\bf 91}, 033111 (2015)

\bibitem{Yoda} Taiki Yoda, Takehito Yokoyama, Shuichi Murakami,  Scientific Reports 5, 12024 (2015).


\end{thebibliography}
\end{document}